\documentclass[preprint,english,aps,reprint,onecolumn,10pt,showkeys]{revtex4-1}
\usepackage[T1]{fontenc}
\usepackage{amsmath,amssymb,amsfonts,hyperref,color,soul}
\usepackage{subscript}
\usepackage{subcaption}
\usepackage{graphicx}
\usepackage{nicefrac}
\begin{document}
\title{
High-precision minmax solution of the two-center Dirac equation}
 \author{Ossama Kullie} 
\affiliation{Theoretical Physics at Institute of Physics, 
Department of Mathematics and Natural Science,
          Heinrich-Plett Str. 40, 34132 Kassel, Germany} 
%
 \thanks{Electronic mail: kullie@uni-kassel.de}
\begin{abstract}
We present a high-precision solution of Dirac equation by numerically 
solving the minmax two-center Dirac equation with the finite element 
method (FEM).
The minmax FEM provide a highly accurate benchmark result for systems 
with light or heavy atomic nuclear charge $Z$.
A result is shown for the molecular ion ${\rm H}_2^+$ and the heavy 
quasi-molecular ion ${\rm Th}_2^{179+}$, with estimated fractional 
uncertainties of  $\sim 10^{-23}$ and $\sim 10^{-21}$, respectively.
The result of the minmax-FEM high-precision  of the solution of the two-center 
Dirac equation, allows solid control over the required accuracy level 
and is promising for the application and extension of our method. 
\end{abstract}
\maketitle
\section{Introduction}\label{Int}
It is well known in quantum chemistry that the Dirac equation is 
subjected to numerical difficulties due to the negative continuum 
of the spectrum.  
This causes a variational instability, which makes the numerical 
computations of one-particle bound states of Dirac equations difficult.
In fact, the Dirac energy functional, which is unbounded from both 
sides is subject to serious implications for variational solutions. 
Variation of the Dirac functional without efficient discretization of 
the negative continuum are therefore, subject to the well known
variational collapse, positronic contamination and the existence of 
spurious states. 
From a numerical viewpoint, the variational collapse
and the existence of spurious states are serious problems.
The minmax formulation of the relativistic one-particle Dirac equation,  
which is used in the present work, Refs. 
\cite{Dolbeault:20002,Dolbeault:2003,Esteban:1999,Griesmer:1999,Talman:1986}, 
has a fundamental property, namely an efficient projection against the 
negative (positronic) continuum. 
{This leads to the minmax energy 
functional, which guarantees that the solution of the Dirac equation 
is restricted to the electronic subspace and is well defined for 
the Coulomb potential with point nuclear charge if 
$Z < Z_{cr}=1/\alpha < 137.036..$, otherwise Dirac operator 
$\widehat{H}_{D}$ is not well defined as a self-adjoint operator} 
\cite{Dolbeault:20002}. 
In the minamx formulation of the Dirac equation, a nonlinear 
dependence on the eigenvalue occurs.
It is, however, even for heavy systems sufficiently weak and does not 
cause a problem in iterative linearized computations of the 
eigenvalues.
Furthermore, in the  non-relativistic limit ($c \rightarrow \infty$) 
the resulting equation goes over directly into the non-relativistic 
Schr\"odinger one, as we will see below.
The spectrum is in accord with the variational characterization of 
the eigenvalues of the Dirac operator based on the minimax principle 
\cite{Kullie:20042,Kullie:20041}, all levels of the computed 
spectrum approach the exact Dirac eigenvalue from above
with the finer approximation of the space (finer subdivisions by 
increasing the grid size), and no indication of spurious energy 
was found.

In the present work, we apply the method developed in earlier studies 
\cite{Kullie:20042,Kullie:20041} and perform a high-precision 
numerical solution of the Dirac equation for an electron in the field 
of two static positive charges. 
We achieved an accuracy of $\sim 10^{-28}$\, atomic units $(au)$ for 
${\rm H}_2^+$ increasing the accuracy achieved in the previous work 
\cite{Kullie:2022} by many orders, and of $10^{-18}$ $au$ for 
${\rm Th}_2^{179+}$, which is much better than our earlier result of 
\cite{Kullie:20041} by many orders as well.
The obtained fractional uncertainty of the relativistic 
{\em shift} is $\sim 10^{-23}$, $\sim 10^{-21}$ for ${\rm H}_2^+$ and 
${\rm Th}_2^{179+}$, respectively.  
The result compares very well with high-precision results recently 
published in the literature, see below.

The remainder of this paper is organized as follows. 
In Sec. \ref{Method}, we present a brief introduction to  
the minimax approach   
and explain the iteration procedure and non-relativistic limit, 
the implementation of which is given in Sec. \ref{Met}.   
To better follow the results and discussion, we briefly
introduce in  Appendix \ref{AA}, some of the theoretical basis 
contained in previous works, in particular references 
\cite{Kullie:20042,Kullie:20041,Kullie:2022}.
In Sec. \ref{RaD} we present our result and discuss the convergence 
and the accuracy of the FEM calculations. 
Finally, a comparison with various results from the literature and 
a conclusion and outlook are presented.
\section{Method}\label{Method}
A solution of the one-particle Dirac equation, a 4-component-spinor 
$\psi$, can be obtained from a stationarity principle of the functional 
$I=\langle\psi|\widehat{H}_D|\psi\rangle-\varepsilon\langle\psi|\psi\rangle$, 
where $\widehat{H}_D, \varepsilon$ are the one-particle Dirac operator 
and energy, respectively.
However, one cannot apply a variational minimum principle as for the 
Schr\"odinger equation, since the spectrum of $\widehat{H}_D$ consists of positive 
(electronic) and negative (positronic) energies.    
The main idea of the minmax method, see \cite{Dolbeault:20002,Talman:1986} 
(and references in \cite{Dolbeault:20002}), is to consider the the subspace 
of electronic states ($F_+$) by projecting out the the subspace of positronic 
states ($F_-$) in a two-step search of extrema, where the sequence of minmax 
level energies is given \cite{Dolbeault:20002} by 
\begin{equation}
\label{eq1} 
\lambda_{k}=\inf_{\stackrel{{\rm dim} G=k}{G 
\mbox{\footnotesize{ subspace of }} F_{+} \,\,\,}} 
\sup_{\stackrel{\psi \in (G\oplus F_{-})}{\psi\neq 0}}
\frac{ \langle \psi \mid \widehat{H}_{D} \mid\psi \rangle}{ \langle  
\psi\mid  \psi\rangle}\ , 
\end{equation}
where $F_{+}\oplus F_{-}$ is an orthogonal decomposition of 
a well-chosen space of smooth square integrable functions and 
$I={\langle \psi \mid \widehat{H}_D \mid\psi \rangle}/{\langle \psi\mid  
\psi \rangle}$ is the energy functional or the Rayleigh quotient, where 
$\psi$ should not be vanishes (hence the vacuum state $\psi=0$ is excluded).
$\psi=(\psi_{1},\psi_{2},\psi_{3},\psi_{4})\equiv (\phi_{+}, \,
\phi_{-})$ is the 4-component relativistic wave function (or short 
4-spinor), with the two spinors  $\phi_{+}=(\psi_{1},\psi_{2})$ and 
$\phi_{-}=(\psi_{3},\psi_{4})$, which are usually called 
upper and lower components of $\psi$ 
and are usually referred to as the large and small component of the 
4-spinor $\psi$. 
For now, we ignore the dependence on the coordinates, which becomes 
clear below.
As already mentioned the minmax functional in eq. \ref{eq1} well 
defined for point nuclear Coulomb potential if 
$Z<1/\alpha<137.036..$), otherwise $\widehat{H}_{D}$ is not well 
defined as a self-adjoint operator (for details on this we kindly 
refer to the review \cite{Smits:2023}), and it has been proven 
\cite{Dolbeault:20002} that the sequence of minmax energies 
$\lambda_{k}$ corresponds to the sequence of positive eigenvalues, 
which represents the electronic part of the total interval 
$(-m c^2, +m c^2)$ of $\widehat{H}_{D}$.
In other words, it guarantees the solution of the Dirac equation in the 
space of the large component $\phi_{+}$, compare eq. 
\ref{eq:integralform} below. 
It is free from spurious states and contaminations that is known from 
the 4-spinor calculations.
The one-particle Dirac eigenvalue equation of the electron in a scalar 
potential $V$, $\widehat{H}_{D}\, \psi= \varepsilon \psi$, can be 
written in the form 
\begin{eqnarray}\label{eq:Diraceq}
 \left(  
\begin{array}{cc} V & \widehat L \\ 
                 \widehat L\ &  V- 2 m_e c^2
\end {array} \right)
\left( \begin{array}{c}
  \phi_{+}\\
  \phi_{-}
\end{array} \right)   
=  \varepsilon  \left( \begin{array}{c}
  \phi_{+}\\
  \phi_{-}
\end{array} \right), 
\end{eqnarray}
where $\widehat L=-i \, c \, \hbar \, {\boldsymbol{\sigma}} \cdot 
\boldsymbol{\nabla}=-i \, c \, \hbar \,\sum_{k=1}^{3}\sigma_k\, 
\partial/\partial x_{k} $, where $x_k$ are the space coordinates 
and $\sigma_k$ are the Pauli matrices.
And because $L$ is a hermitian operator $\langle\phi|L^{\dagger}=
L|\phi\rangle$, we ignore the ${\dagger}$ sign. 
By  eliminating  the small component $\phi_{-}$ from eq. 
(\ref{eq:Diraceq})  one obtains 
\begin{equation} 
\label{eq:LdaggerL}
\widehat L\left(\frac{\widehat L \, 
\phi_{+}}{\varepsilon + 2 m c^2  -V}\right) 
= (\varepsilon - V)\,\phi_{+}\ ,
\end{equation} 
$\varepsilon$ is the eigenenergy that in the non-relativistic limit 
corresponds to the eigenenergy of the Schr\"odinger equation.  
{Eq.} (\ref{eq:LdaggerL}) can be now transformed into the minmax 
integral ``weak'' form, which offers a good efficiency for FEM with 
large finite-element basis set, leading to the equation, 
\begin{equation}\label{eq:integralform}
\int  \frac{ c^{-2}| L\phi_{+}|^2}{2m_{e} + (\varepsilon -V)/ c^2} dr^3 = 
\int (\varepsilon - V)\,|\phi_{+}|^2dr^3\ .
\end{equation}
The minmax principle now guarantees the minimum of the energy value 
$\varepsilon$ as mentioned above \cite{Dolbeault:20002}.
Obviously, the two equations (\ref{eq:LdaggerL}), (\ref{eq:integralform}) 
bear similarities to their non-relativistic counterparts, the 
Schr\"odinger equation and its integral ``weak'' form since 
$\lim_{c\to \infty} \nicefrac{(\varepsilon-V)}{c{^2}}=0$, in which the 
two components $\psi_{1}, \psi_{2}$ of the $\phi_+$ transform into their 
non-relativistic counterparts $\phi_{\uparrow},\phi_{\downarrow}$ 
(they are degenerate in the absence of a magnetic field interaction).  
However, eqs. (\ref{eq:LdaggerL}), (\ref{eq:integralform}) are nonlinear 
in the eigenvalue $\varepsilon$ and therefore have to be solved by 
iteration.
The nonlinearity is weak hence the left-hand side of eq. 
(\ref{eq:integralform}) is expanded in a series \cite{Dolbeault:20002} 
as follows. 
With an approximate value $\varepsilon_0$ of an eigenvalue $\varepsilon$, 
the iteration procedure at iteration $j+1$ $(j=0,...,j_{\rm max})$ 
consists of expanding the left-hand side as \cite{Kullie:20042},
\begin{eqnarray}\label{eq:E0jjexpansion}
&&\int\frac{|L\phi_{+}|^2}{\varepsilon^{j}+ 2 m_e c^2 -V} dr^3=
\int\frac{|L\phi_{+}|^2}{g(\varepsilon_0)}dr^3
+\sum_{k=1}^{k_{\rm max}} (-{\Delta \varepsilon^{j}})^{k} 
\int\frac{|L\phi_{+}|^2}{g(\varepsilon_0)^{k+1}} dr^3\ ,
\end{eqnarray}
with $g(\varepsilon_0)=\varepsilon_0 + 2 m_e c^2 -V$ and 
$\Delta  \varepsilon^{j} = \varepsilon^{j} - \varepsilon_0$. 
The first term, the matrix elements, on the right-hand side of eq. 
\ref{eq:E0jjexpansion} is computed once and stored for the next 
iterations.  
The second term is updated on each iteration $j\ge1$, requiring only 
multiplication operations and a sum over $k$.  
The series converges rapidly, where $k_{\rm max}=3-9$ is sufficient for 
$Z=1-90$ since is it easy to guess an approximate value $\varepsilon_0$ 
(e.g. from non-relativistic values). 
The matrix equation is solved by an iterative method with a Cholesky 
decomposition \cite{Heinemann:1987}, which is the heaviest part of the 
computation. 
More details can be found in \cite{Kullie:20042,Heinemann:1987}. 

In our FEM approach, as we will see in the next section, we perform 
the computation for a series of successive grids and the approximate 
solution of one grid is used to start the next finer grid. 
{Compared to the 4-spinor formulation, the minmax formulation exhibits  
major advantages: only $\nicefrac{1}{3}$ of the matrix elements of the
4-spinor formulation, have to be computed and the vector iteration 
requires a factor $4$ less operations} \cite{Kullie:20041,Kullie:20042}.
The reduced size of the problem enhances the computational performance 
considerably, as we will see in the present work. 

We have seen that in the limit ($c\to \infty$), eq. (\ref{eq:integralform}) 
transforms directly to the non-relativistic the integral ``weak'' form 
of the Schr\"odinger equation and therefore is expected to exhibit 
similar properties to the non-relativistic case.
The non-relativistic eigenvalue is calculated by setting $c$ to a large 
number (in the present work $c> 10^{+15},\, \alpha^{2} < 10^{-30}$, 
the uncertainty is of the order $(\delta \alpha)^{2} \sim 10^{-31}$, 
see further below). 
This has an important advantage that the the relativistic 
{\em shift} is extracted with a better accuracy than the directly  
calculated values, due to error cancellation (see Sec.\, \ref{RaD}).
It is known in computational chemistry by the acronym counterpoise.  
\subsection{Solution of the two-center Dirac equation with FEM}\label{Met}
The Dirac Hamiltonian in $au$ for a single-particle (of mass 
$m=1$) in a two-center potential $V$ is 
\begin{eqnarray}\label{eq:hhD} 
{H}_D = c \,\boldsymbol{\alpha}
               \cdot {\bf{p}} + c^2 {\beta} +V\ , \\ \nonumber
\mbox{} V = - \sum_{l=1}^2 \frac{Z_l}{|{\bf r} - {\bf R}_l|}\ .
\end{eqnarray}
where ${\boldsymbol \alpha}$ and $\beta$ the usual Dirac matrices, 
${\bf{p}}$ is the momentum.
$Z_{1}, Z_{2}$ are the charges of the two nuclei in units of the 
elementary charge, ${\bf r}$ is the position of the electron, and 
${\bf R}_l$ are the positions of the nuclei. 
In the two-center case one has axial symmetry around the internuclear 
axis (the $z$-axis), which suggests the use of prolate spheroidal 
(elliptic spheroidal) coordinates $\xi $, $\eta $ and $\varphi$.
A further singular coordinate transformation ($\xi(s),\eta(t)$), 
see eq. (\ref{eq:transf}), is required to deal with the Coulomb 
singularity of point nucleus. 
It guarantees a high order of convergence that allows full use of the 
higher FEM (approximation) order $p$, as it provides 
an efficient description of the singularity of the wave function in the 
vicinity of the nuclei \cite{Kullie:2001,Yang:1993,Yang:19911}.

Due to axial symmetry the angular dependence is separated analytically 
by the ansatz:
\begin{eqnarray}\label{ANSATZ} 
\psi&=&
  \left(\begin{array}{c}
  \phi_{+}(s,t,\varphi)\\
  \phi_{-}(s,t,\varphi)
\end{array}\right)\!=\!       
\left( \begin{array}{c}
  \phi^1(s,t) \cdot e^{i(j_{z}-1/2) \cdot \varphi}\\
  \phi^2(s,t) \cdot e^{i(j_{z}+1/2) \cdot \varphi}\\
  i\phi^3(s,t) \cdot e^{i(j_{z}-1/2) \cdot \varphi}\\
  i\phi^4(s,t) \cdot e^{i(j_{z}+1/2) \cdot \varphi}\\
\end{array}\right)\ .
\end{eqnarray}
the $z$-component of the total angular momentum $j_z$, is a good 
quantum number and the relativistic wave function $\psi$ is an 
eigenstate of the total angular momentum and its $z$-component $j_z$. 
With the Hamiltonian $H_{D}$ of eq. \ref{eq:hhD}, and eqs 
(\ref{eq:Diraceq})-(\ref{ANSATZ}) the two dimensional Dirac equation 
is transformed into eigenvalue matrix equation, which is solved 
numerically by FEM (and iteratively), as already mentioned 
\cite{Kullie:2022}.
The 2-dimensional domain $(s,t)$ is subdivided into FEM triangular 
elements and the components $\phi^k(s,t)$ are approximated using 
global functions and a complete polynomials of an order $p$ in $s,t$  
\footnote{A two-variable, p-order, complete polynomial is 
defined by: \\ ${\cal{P}}(s,t)=a_{00}+ a_{10}\,s+ a_{01}\,t + a_{11}\,s\,t 
+ a_{20}\,s^2 + a_{02}\,t^2 +\dots + 
a_{p0}\,s^p + a_{0p}\,t^p$}, 
see Appendix Sec \ref{glob1}.
\section{Result and Discussion}\label{RaD}
\begin{table}[t]
\scriptsize
  \begin{tabular}{c|l|l|l} \hline \hline  
 $N_{e}/N$& \hspace{1cm}Relativistic, $E_{\rm rel}$  & 
  \hspace{0.5cm} Non-relativistic, $E_{\rm nrel}$ & \hspace{0.3cm} Rel. {\em shift}
  $\Delta E_{\rm rel}$ $(10^{-6})$ \\ \hline
 \tiny 72/3721   & {\bf-1.10264158}076265658336304 & {\bf-1.102634214}22500083644351 & {\bf-7.3665376}55746919537216 \\
 \tiny 128/6561  & {\bf-1.10264158103}129731540804 & {\bf-1.10263421449}366618700516 & {\bf-7.366537631}128402879085 \\
 \tiny 200/10201 & {\bf-1.1026415810325}4876503352 & {\bf-1.1026342144949}1805349370 & {\bf-7.3665376307}11539825949 \\
 \tiny 288/14641 & {\bf-1.10264158103257}440524021 & {\bf-1.10263421449494}370230232 & {\bf-7.366537630702}937883947 \\
 \tiny 392/19881 & {\bf-1.102641581032577}01626914 & {\bf-1.102634214494946}31361918 & {\bf-7.3665376307026}49958534 \\
 \tiny 512/25921 & {\bf-1.10264158103257716}018477 & {\bf-1.1026342144949464}5757546 & {\bf-7.366537630702609}305534 \\
 \tiny 648/32761 & {\bf-1.10264158103257716}288008 & {\bf-1.10263421449494646}027071 & {\bf-7.366537630702609}365171 \\
 \tiny 800/40401 & {\bf-1.10264158103257716}398460 & {\bf-1.102634214494946461}37541 & {\bf-7.3665376307026091}81867 \\
 \tiny 968/48841 & {\bf-1.102641581032577164}09888 & {\bf-1.102634214494946461}48972 & {\bf-7.36653763070260915}7530 \\
 \tiny 1152/58081& {\bf-1.10264158103257716411}642 & {\bf-1.10263421449494646150}726 & {\bf-7.366537630702609156}421 \\
 \tiny 1352/68121& {\bf-1.10264158103257716411}686 & {\bf-1.10263421449494646150}770 & {\bf-7.366537630702609156}250 \\
 \tiny 1568/78961& {\bf-1.10264158103257716411}800 & {\bf-1.102634214494946461508}84 & {\bf-7.3665376307026091560}39 \\
 \tiny 1800/90601& {\bf-1.1026415810325771641181}1 & {\bf-1.1026342144949464615089}5 & {\bf-7.36653763070260915605}5 \\
${\rm extp}^{1}$ & {\bf-1.1026415810325771641181}4 & {\bf-1.1026342144949464615089}8 & {\bf-7.3665376307026091560584}\\ 
\hline \hline 
\end{tabular}
\caption{\scriptsize 
Energies of {${\rm H}_2^+$} at $R=2$. All values in atomic units.
The calculations utilize $\nu=8$ and $D_{\rm max}=40$. $N_{e}, N$ are 
the numbers of the elements and grid points respectively.  
Last digit is rounded. 
Superscript $^{1}$ indicates values extrapolated over the sequence 
$N_{e}$.  Bold digits are significant.}\label{tab:D40Nu8} 
\end{table}
In the present work, we compute the ground state energy 
$\varepsilon_{1(1/2)g}$, i. e. with $j_z=1/2$ and gerade symmetry $g$, 
for two molecular ions ${\rm H}_2^+$ and ${\rm Th}_2^{179+}$. 
The notation of the corresponding non-relativistic state is 
${1\sigma_g}$. 
We abbreviate the notation by the short-hand $E_{\rm rel},\,
E_{\rm nrel}$, respectively. 
In our calculation, unless otherwise specified, we use atomic units 
and the CODATA 2018 value of $c= \alpha^{-1} = 137.035999084$ 
\cite{Tiesinga:2021}. 

In the following, the notation $p$ denotes the order of a two-variable, 
complete polynomial (FEM-approximation), and $\nu$ refers to the values  
generating the sum of the singular coordinate transformation, 
see eq. \ref{eq:transf}.
Further notations will be explained in the text. 
In the presented calculations, we use the FEM polynomial order 
$p=10$, which approximately guarantees a convergence order 
$\sim 10$ in the calculation by suitable $\nu$ value of the singular 
coordinate transformation of eq. \ref{eq:transf}, as we will see below. 
We run test calculations for different values of $\nu$ and size of the 
domain in order to check and reach the highest convergence. 
The size is defined by the size of the largest ellipse  
$\xi=\xi_{\rm max}=const.$ containing the grid elements. 
The size of the grid can alternatively be defined by 
$D_{\rm max}(\xi_{\rm max})$, defined as the distance (perpendicular 
to the connecting line of the two centers) between one of the centers 
to a point on the outermost ellipse $\xi_{max}$\cite{Kullie:2003}. 
It was found that the optimal $D_{\rm max}$ values (given in atomic units) 
of $\sim 40$ for ${\rm H}_2^+$ and $\sim 0.35$ for ${\rm Th}_2^{179+}$.  
However, as shown below, the precise value of $D_{\rm max}$ is not crucial. 
It should be compared to 
the internuclear distance, $R=2$, $R=2/90$ for the ${\rm H}_2^+$ and 
${\rm Th}_2^{179+}$, respectively.
$R=2$ is the approximate equilibrium bond length of the ${\rm H}_2^+$ 
molecule, whereas $R=2/90$ is scaled over the atomic number $Z=90$ 
for ${\rm Th}_2^{179+}$.

In  table \ref{tab:D40Nu8}, we show the calculated energies for the 
ground state of the ${\rm H}_2^+$.  
In the first column the element and point number of the grids are given. 
The relativistic $E_{\rm rel}$ and non-relativistic $E_{\rm nrel}$ 
energy values are given in the columns $2, 3$. 
As seen in the table the accuracy increased systematically for both 
values of $E_{\rm rel}$ and $E_{\rm nrel}$.
The relativistic {\em shift} is given in column 4, which is calculated 
by the simple relation $\Delta E =E_{\rm rel}-E_{\rm nrel}$. 
The absolute uncertainty of the relativistic {\em shift} is better than 
the calculated values of the respective energies, benefiting from error 
compensation \cite{Kullie:20041} by considering the non-relativistic 
energy $E_{\rm nrel}$ for the same grid and parameters. 
In addition, table \ref{tab:D40Nu8} (last row) shows the extrapolated 
values \cite{Kullie:20042,Kullie:20041,Kullie:1999} over 
grid elements (or grid points as both sequences scale similar). 
For clarity, bold digits indicate significant digits.
As seen in table \ref{tab:D40Nu8}, the accuracy increases with increasing 
number of grid points (finer subdivision), and that 
the convergence to the exact (lower energy value) is from above, not 
only for the non-relativistic but also for the (Dirac) 
relativistic energy value, which is not surprising (see Sec. 
\ref{Method}), and is a major advantage of minmax approach, which 
effectively projects the problem onto the electronic states.
\begin{figure}
\centering
\begin{subfigure}{0.4\textwidth}
    \includegraphics[width=\textwidth]{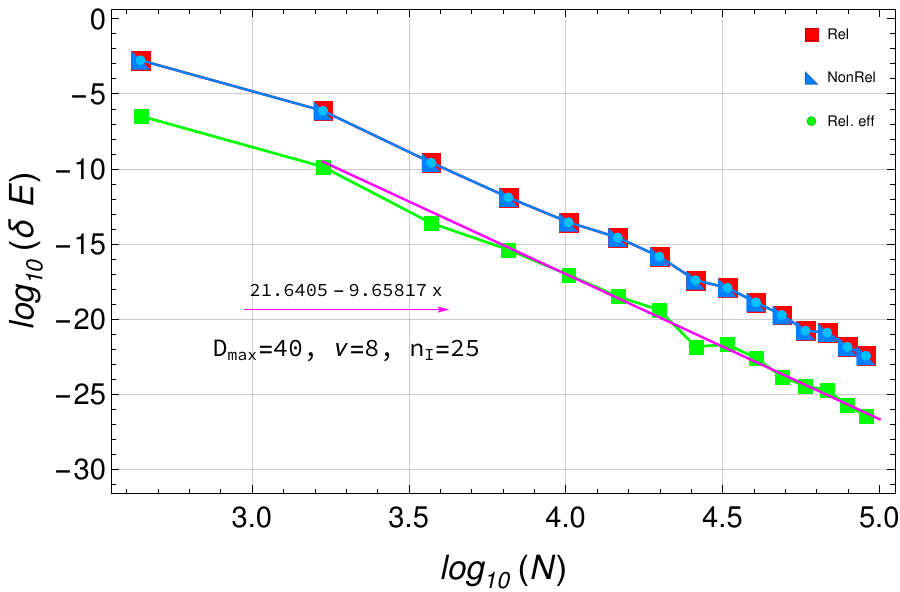}
    \caption{$D_{\rm max}=40\, au$, $\nu=8$, good convergence.}
    \label{fig:1a}
\end{subfigure}
\hspace{1.cm}
\begin{subfigure}{0.4\textwidth}
    \includegraphics[width=\textwidth]{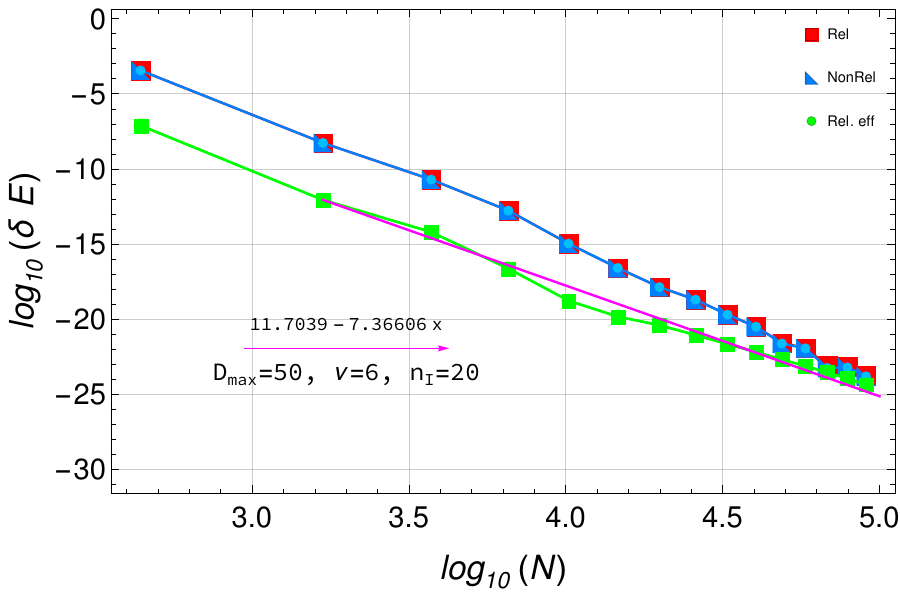}
    \caption{$D_{\rm max}=50\, au$, $\nu=6$, moderate convergence.} 
    \label{fig:1b}
\end{subfigure}     
 \caption{\label{fig:H2nu} \scriptsize (Color online) 
 Convergence behavior of the relativistic (red square) and 
 non-relativistic (blue, left triangle) energies and of the 
 relativistic {\em shift} (green circle) for  ${\rm H}_2^+$, 
 as a function of the number of grid points $N$. 
 $\delta E=|E^{\rm extp}-E{(\rm N)}|$ (see table  \ref{tab:D40Nu8}, 
 where the first two points are omitted).  
 Two cases are shown.
 (a): $D_{\rm max}=40\, au$, $\nu=8$, highest possible convergence 
 order of polynomial approximation $p=10$ (see table \ref{tab:D40Nu8}). 
 (b): $D_{\rm max}=50\, au$, $\nu=6$, a moderate convergence order, which 
 is considerbly below the FEM order of p=10 (values are not shown).}
\end{figure}

In  Fig. \ref{fig:H2nu}, we present a log-log plot of the errors 
$\delta E(N)$ of the energies and of the relativistic {\em shift}, 
with respect to the extrapolated value (the last row in table 
\ref{tab:D40Nu8}). 
As can be seen from the red and blue colored lines (points) in  
Fig. \ref{fig:H2nu}, the convergence rate in the (relativistic) minmax 
formulation is close to that of the non-relativistic Schr\"odinger 
formulation \cite{Zhang:2004}.
Concerning the convergence, a high value of $\nu$ (e.g. $\nu=8$) in 
the coordinate transformation eq \ref{eq:transf} is needed (especially 
for grids with a large number of points, compare Fig. \ref{fig:H2nu}), 
which in turn enables a higher convergence order for the energy and 
the full utilization of a higher FEM order approximation $p=10$
\cite{Kullie:2001,Kullie:2003,Kullie:20041}.

For a larger $\nu$ value (Fig. \ref{fig:1a}), the wave 
function is better approximated near the two centers, and the error 
of the relativistic singularity eq. \ref{glob2} becomes significantly 
small also for the dense grids.
Here, a suitable distribution of the grid points between the inner and 
outer regions over the domain is thereby achieved. 
This optimizes error compensation and achieves better accuracy in the 
relativistic {\em shift} (green colored).
As an illustration, we performed calculations for the same grids 
as in table \ref{tab:D40Nu8} but with a lower value $\nu=6$ 
and present the corresponding log-log plot in Fig. 
\ref{fig:1b}.
From the linear fit on the relativistic {\em shift} shown in the Fig. 
\ref{fig:1a}, using $\nu=8$, one finds a convergence order 
$q\approx 9.7$, which is close to the FEM order of $p=10$, unlike 
in the Fig. \ref{fig:1b}, where $q\approx 7.4$ is considerably smaller than 
the FEM order $p=10$, especially an optimal cancellation of the errors 
is not achieved for larger grids.   
As seen in Fig. \ref{fig:H2nu}, the estimated uncertainties of the 
computed energies are  $\sim 10^{-23}$, and for the relativistic 
{\em shift} is $\sim 10^{-28}$ or a fractional uncertainty of 
$\sim 10^{-23}$ in the relativistic {\em shift} (see also below 
table \ref{tab:comp}).
\begin{table}[h]
\scriptsize
  \begin{tabular}{ccc} \hline \hline
 \(\begin{array}{c|c|c}
 D_{\rm max} & {\rm value\ for\ densest\ grid} & {\rm extrapolated\, value}
 \\\hline
 30 & -7.3665376307026091560591\, & {\bf -7.36653763070260915605}76 \\
 40 & -7.3665376307026091560546\, & {\bf -7.366537630702609156058}3\\
 50 & -7.3665376307026091560635\, & {\bf -7.366537630702609156058}1\\
 60 & -7.3665376307026091560496\, & {\bf -7.3665376307026091560}250\\
 \hline \hline
\end{array}\)
\end{tabular}
\caption{\scriptsize
Dependence of the relativistic {\em shift} $\Delta E_{\rm rel}$ (in $10^{-6}$ 
atomic units) at $R=2$ on the domain size $D_{\rm max}$. 
The result is for the densest grid with $1800/90601$ 
elements/points. 
All evaluations were performed with $\nu=8$. Bold digits are significant}
\label{tab:relef1}
\end{table}

Table \ref{tab:relef1} shows the dependence of the relativistic 
{\em shift} on the domain size $D_{max}$ and makes it clear that the 
exact value of the domain size does not matter.
Table \ref{tab:relef1} indicates that the optimal value is in 
region $D_{\rm max}\sim  40$.  
In fact, a moderate variation of $D_{\rm max}$ mainly affects the outer 
region; therefore, due to the error cancellation, the relativistic 
{\em shift} is not sensitive to $D_{\rm max}$.
\begin{table}[h]
\scriptsize
  \begin{tabular}{c|l|l|l} \hline  \hline
 $N_{e}/N$& \hspace{1cm}Relativistic, $E_{\rm rel}$  & 
  \hspace{0.5cm} Non-relativistic, $E_{\rm nrel}$ & \hspace{0.3cm} Rel. {\em shift} $\Delta E_{\rm rel}$ 
  \\ \hline
 \tiny 72/3721 &   {\bf -9504.75662}77711897646180& {\bf -8931.337}058411524371542 & {\bf -573.4195}693596653930759 \\
 \tiny 128/6561&   {\bf -9504.756648}3577412426133& {\bf -8931.337137}096470274648 & {\bf -573.419511}2612709679648 \\
 \tiny 200/1020&   {\bf -9504.756648}4301451994401& {\bf -8931.337137}399365527444 & {\bf -573.4195110}307796719956 \\
 \tiny 288/14641&  {\bf -9504.75664843}3886680448 & {\bf -8931.33713740}8143475088 & {\bf -573.41951102}57432053607 \\
 \tiny 392/19881&  {\bf -9504.756648434}005781759 & {\bf -8931.3371374090}57756356 & {\bf -573.41951102494}80254030 \\
 \tiny 512/25921&  {\bf -9504.75664843400}8746274 & {\bf -8931.33713740906}3219487 & {\bf -573.41951102494}55267868 \\
 \tiny 648/32761&  {\bf -9504.7566484340094}21628 & {\bf -8931.337137409066}302506 & {\bf -573.4195110249431}191218 \\
 \tiny 800/40401&  {\bf -9504.7566484340094}83622 & {\bf -8931.337137409066}299523 & {\bf -573.4195110249431}840987 \\
 \tiny 968/48841&  {\bf -9504.7566484340094}96581 & {\bf -8931.33713740906633}5431 & {\bf -573.41951102494316}11496 \\
 \tiny 1152/5808&  {\bf -9504.7566484340094}99723 & {\bf -8931.33713740906633}7662 & {\bf -573.419511024943162}0606 \\
 \tiny 1352/68121& {\bf -9504.756648434009500}459 & {\bf -8931.337137409066338}069 & {\bf -573.419511024943162}3896 \\
 \tiny 1568/78961& {\bf -9504.756648434009500}656 & {\bf -8931.337137409066338}170 & {\bf -573.419511024943162}4852 \\
 \tiny 1800/90601& {\bf -9504.7566484340095007}23 & {\bf -8931.3371374090663382}16 & {\bf -573.4195110249431625}066 \\
 ${\rm extp}^{1}$  & {\bf -9504.7566484340095007}48 & {\bf -8931.3371374090663382}35 
 & {\bf -573.41951102494316251}4 \\ 
\hline  \hline
\end{tabular}  
\caption{\scriptsize
Energies of {${\rm Th}_{2}^{179+}$} at $R=2/90$. 
All values in atomic units.
The calculations utilize $\nu=10$ and $D_{\rm max}=0.35\, au$. $N_{e}, 
N$ are the numbers of the elements and grid points respectively.  
Last digit is rounded. 
Superscript $^{1}$ and bold digits as in table \ref{tab:D40Nu8}.}
\label{tab:D035Nu10}  
\end{table}

\begin{figure}[t]
\centering
\begin{subfigure}{0.4\textwidth}
    \includegraphics[width=\textwidth]{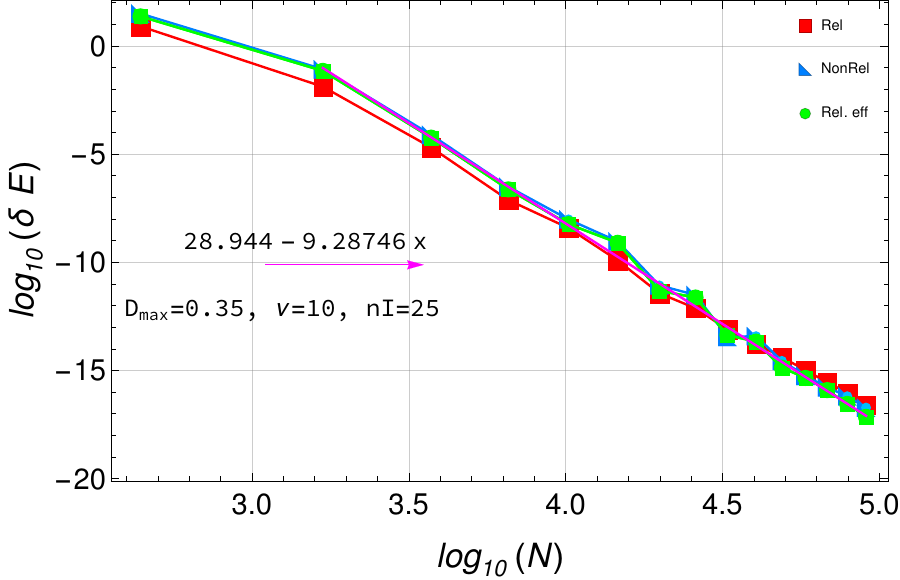}
    \caption{With $D_{\rm max}=0.35, \nu=10$ and integration points $n_I=25$.}
    \label{fig:2a}
\end{subfigure}
\hspace{1.cm}
\begin{subfigure}{0.4\textwidth}
    \includegraphics[width=\textwidth]{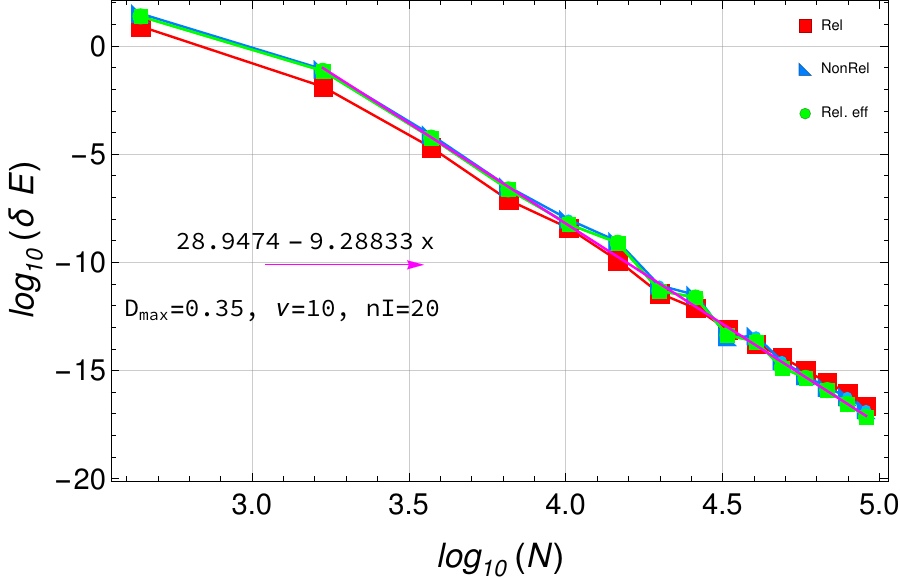}
    \caption{Same as (a) but with integration points $n_I=20$.} 
    \label{fig:2b}
\end{subfigure}    
 \caption{\label{fig:Th2nu} 
  \scriptsize (Color online) 
 Same as in Fig. \ref{fig:H2nu}, for the quasi-molecular ion ${\rm Th}_2^{179+}$.
 (a) and (b) are calculated with $D_{\rm max}=0.35, \nu=10$ and FEM polynomial 
 approximation order $p=10$. 
 For (a) compare table \ref{tab:D035Nu10} (whrer the first two points are omitted).}  
 \end{figure}

For systems with high atomic nuclear charge $Z$, such as our calculated  
system ${\rm Th}_2^{179+}$, it has already been shown \cite{Kullie:20042}, 
\cite{Kullie:20041} that the minmax formulation behaves much 
better than the 4-spinor formulation. 
The achieved accuracy of the calculations is significantly higher, 
while the computational effort is lower.
In table \ref{tab:D035Nu10}, we report our result for ${\rm Th}_2^{179+}$ 
for the same grid sequence used in table \ref{tab:D40Nu8} for ${\rm H}_2^{+}$. 
For ${\rm Th}_2^{179+}$, a small domain size $D_{\rm max}=0.35\, au$ and $\nu=10$ 
are used. 
Similar to ${\rm H}_2^{+}$, despite the strong singularity, one finds a convergence 
from above towards the exact values of the non-relativistic and the 
Dirac relativistic energies, which demonstrates the power of the FEM-minmax 
method.  
The absolute accuracy is reduced compared to the ${\rm H}_2^{+}$, this is 
because for high-$Z$ the singular behavior is much stronger and a higher 
density of the grid points needed for the regularization of the 
singularity near the nucleus.
This in turn affects the approximation error (dilutes the point 
density) at large distance from the centers.
The value of $\nu=10$ is large but still the singularity error 
outweighs the convergence order.
This can be seen in Fig. \ref{fig:Th2nu}, where similar to Fig.  
\ref{fig:H2nu}, we show a log-log plot of the errors $\delta E(N)$ of 
the energies and the relativistic {\em shift} relative to the 
extrapolated value (the last row in table \ref{tab:D035Nu10}). 
As seen in Fig. \ref{fig:Th2nu}, the error in the relativistic 
{\em shift} is only slightly better than in the energy values, 
unlike what we found in ${\rm H}_2^{+}$, since the singularity error hinders an 
efficient cancellation of the smooth FEM (or the FEM-approximation) 
error at the outer region far from the nucleus. 
Nevertheless, a convergence order $q\approx 9.3$ is reached, it is 
below the FEM order of $10$, although it is not far from that of 
${\rm H}_2^{+}$, as seen from the linear fit in Figs. \ref{fig:H2nu}, 
\ref{fig:Th2nu}.

To test the accuracy and the convergence of the result by checking the 
accuracy of the matrix element, we achieved similar calculations but 
with smaller integration point per (triangular FEM) element $n_I=20$  
instead of $n_I=25$ per element.
The result is log-log plotted in Fig. \ref{fig:2b}, in  
a similar way to Fig. \ref{fig:2a}. 
The two results with $n_I=25, n_I=20$ show a similar behavior. 
They illustrate and confirm that the (lower) convergence behavior is  
caused by the singularity near the centers at such a high $Z$.
A similar test is done for ${\rm H}_2^+$, it was found that the effect of the 
integration point between $n_I=20$ and $n_I=25$, is of the order 
$\delta\varepsilon(n_I=20\rightarrow n_I=25) \sim 10^{-29}$. 
And using $n_I>25$ does not improve the accuracy.

The value of $D_{\rm max}$ is also not crucial for the calculation 
or ${\rm Th}_2^{179+}$, but the error cancellation 
($E_{\rm rel}-E_{\rm nrel}$) is more sensitive to $D_{\rm max}$ than for 
${\rm H}_2^+$. 
It turns out that a range around $D_{\rm max}\sim 0.35$ is optimal, 
as can be seen in Fig. \ref{fig:Th2nu}, in which 
an optimal balance is achieved between short-range error 
(singularity error) and long-range error, which is on the order 
of the FEM approximation. 

We checked this behavior for different $D_{max}=0.25-0.50$, 
the result is shown in table \ref{tab:Fres}. 
A look at the table shows that the unbalanced distribution of 
the grid points between inner and outer regions causes some 
oscillatory behavior in the relativistic effect, see also below.
From Fig. \ref{fig:Th2nu} and table \ref{tab:D035Nu10} it can be seen  
that the uncertainties in the relativistic energy are estimated at 
$\sim10^{-18}$, or a fractional uncertainty of $ 10^{-21}$ in the 
relativistic {\em shift}, which is two orders of magnitude worse 
than for ${\rm H}_2^+$.
\begin{table}[t]
\scriptsize
\begin{tabular}{lll} \hline\hline
$D_{max}$ & \hspace{2cm} $E_{\rm rel}$ \\ \hline 
0.300& -9504.7566484340095007351 \\
0.325& -9504.7566484340095007376 \\
0.335& -9504.7566484340095007368 \\
0.350& -9504.7566484340095007373 \\
0.365& -9504.7566484340095007383 \\
0.375& -9504.7566484340095007387 \\
0.400& -9504.7566484340095007371 \\
\hline\hline
\end{tabular}
\caption{\scriptsize Scatter of the extrapolated values $E_{\rm rel}$
as a function of $D_{max}$ around the value given in Table 
\ref{tab:comp}.
A lower limit of $E_{\rm rel}$ is  $-9504.75664843400950074$,
which is slightly shallower than the extrapolated value given in table 
\ref{tab:D035Nu10}.\label{tab:Fres}}
\end{table}

Finally, we note that for an effective error cancellation, we use the 
same $\nu$ of the singular coordinate transformation for the 
non-relativistic calculation, although the smallest value of the 
singular coordinate transformation $\nu=2$ for non-relativistic 
calculations is sufficient.
Using $\nu= 2$ does not condense the points in the inner region, which 
means that the density of the points in the outer region (far from the 
core) are not diluted, resulting in better accuracy than with a higher 
$\nu$ value for the same grid points in the non-relativistic 
calculations.    
By adding the relativistic {\em shift} to the non-relativistic values 
calculated with $\nu=2$, one obtains more accurate relativistic energy 
value, which is done in table \ref{tab:comp} that shows our final 
result with a comparison to a recent available result from the 
literature \cite{Nogueira:2023}.  
However, due to the extrapolation the values for different $D_{max}$ 
scatter and a limited gain in the accuracy is reached. 
Table \ref{tab:Fres} shows this behavior and demonstrate the 
reliability of our final result for ${\rm Th}_2^{179+}$ in table 
\ref{tab:comp}.
\begin{table}[h]
\scriptsize
\begin{tabular}{lll} \hline\hline
 & \hspace{2cm}${\rm H}_2^+$ & \hspace{1.3cm}${\rm Th}_2^{179+}$\\ \hline 
Rel. eff. & -0.0000073665376307026091560584  & -573.4195110249431625138\\
Nrel.     & -1.10263421449494646150896894154 & -8931.3371374090663382226\\
Rel.      & -1.10264158103257716411812499995 & -9504.756648434009500737\\
\cite{Nogueira:2023} 
      & -1.102641581032577164118124999957656 & -9504.756648434009500732\\
\hline\hline
\end{tabular}
\caption{\scriptsize Final result and a comparison with the recent 
available result from the literature. 
\label{tab:comp}}
\end{table}

Looking at table \ref{tab:comp} on finds an excellent agreement with 
the result of Nogueira et. al. \cite{Nogueira:2023}. 
The discrepancies are $\sim  7. 10^{-30}, 5. \, 10^{-18}$ (where the 
last digits are rounded) for ${\rm H}_2^+$, ${\rm Th}_2^{179+}$, respectively. 
We have to add that the extrapolated value $E_{\rm nrel}$ in table 
\ref{tab:comp} was obtained with $\nu=2$ and $D_{max}=80, 0.80\, au$ 
for ${\rm H}_2^+$ and ${\rm Th}_2^{179+}$, respectively, and the 
(extrapolated) relativistic {\em shift} was obtained by using the 
convergence orders, $q=9.7, 9.3$ as shown in Figs \ref{fig:H2nu}, 
\ref{fig:Th2nu}, respectively.  
Where $q$ is the leading order of the polynomial or rational 
(function) approximation of the error.
It changes slightly by changes of the parameters, where in general the  
discrepancy scatters around $\lesssim 10^{-29}$, $\lesssim 10^{-18}$ 
or fractional uncertainties of $\sim 10^{-23}, \sim 10^{-21}$ for ${\rm H}_2^+$, 
${\rm Th}_2^{179+}$, respectively. 
This sets the limits of accuracy in our result, considering the various 
parameters and highest grid points used in table \ref{tab:D40Nu8}, 
\ref{tab:D035Nu10}. 
One notices that the extrapolated values of $E_{\rm rel}^{extp},\, 
E_{\rm nrel}^{extp}$ in tables \ref{tab:D40Nu8}, \ref{tab:D035Nu10} 
(in the second and third columns) are slightly lower than the value 
given in table \ref{tab:comp}.
That is because the extrapolation of $E_{nrel},\, E_{rel}$ usually 
yields a lower value than the ``exact'' one, but for the relativistic 
effect a correction in the opposite direction (counterpoise effect) is  
achieved. 
And usually the value {$E^{extp}_{\rm rel\, shift}$ is better 
than the difference $E_{\rm rel}^{extp}-E_{\rm nrel}^{extp}$.  
While in the first case the extrapolation is performed once, in the 
second case the extrapolated result of $E_{\rm rel}^{extp}, 
E_{nrel}^{extp}$ could occur in opposite directions. 
Considering the comparison to the work of Nogueira et. al. 
\cite{Nogueira:2023}, our FEM minmax methods behaves better in 
the relativistic domain, because of two aspects. 
First, the minmax guarantees convergence from above, and second, it is  
free from spurious states and thus free from contamination (see also 
Fillion et al.  \cite{Fillion:2012}), whereas the authors of 
\cite{Nogueira:2023} have reported such states in the calculations 
for $Z=90$. 
This never has been detected in our FEM calculation so far for $Z=90$, 
see ref. \cite{Zhang:2004}.
In our calculation, we find $2N$ ($N$ is the grid points number) 
positive eigenvalues corresponds to the large (2-component) spinor 
$\phi_+$ of the Dirac wave function, see Secs. \ref{Method}, \ref{Met}. 
Taking these two aspects into account, we think that our final result 
in table \ref{tab:comp} for the highly relativistic ${\rm Th}_2^{179+}$ 
is slightly better than the values of \cite{Nogueira:2023}, although for 
${\rm H}_2^+$ the value of \cite{Nogueira:2023} is given with more 
significant digits than our value. 
\begin{table}[h]
\scriptsize
\begin{tabular}{lll}
\hline\hline
Reference & \hspace{1cm}${\rm H}_2^+$ & \hspace{1.cm}${\rm Th}_2^{179+}$
\\ \hline 
 Kullie et al. \cite{Kullie:2022}, \cite{Kullie:20041} & { -1.10264158103360758005}$^{a}$ &
 {-9504.7567469606}$^{a}$ \\ 
 Mironova et al. 2015 \cite{Mironova:2015}        & { -1.102641581033}0 & 
 {-9504.756746927}$^{a}$ \\
 Tupitsyn et al. 2014 \cite{Tupitsyn:2014}        & { -1.1026415810330}$^{a}$ & 
 {-9504.756746927}$^{a}$ \\
 Fillion-Gourdeau et al. 2012 \cite{Fillion:2012} & { -1.102641580782}$^{b,d}$ &
 {-9504.698874401}$^{b,d}$ \\
 Artemyev et al. 2010 \cite{Artemyev:2010}    &  { -1.1026409}$^{c}$& {-9504.752}$^{c}$\\
 Ishikawa et al. 2008 \cite{Ishikawa:2008}        & { -1.102641581033}598$^{a}$  &\\
 Parpia and Mohanty 1995 \cite{Paripa:1995} & {-1.10264158}01$^{a}$ & {-9504.756696}$^{a}$\\
 Rutkowski 1999 \cite{Rutkowski:1999}           &         & {-9504.7567151}$^{a}$ \\
\hline \hline
\end{tabular}  
\caption{\scriptsize
Comparison with values for the literature with different $\alpha$ values, 
for the ground-state energy of the ${\rm H}_2^{+}$, ${\rm Th}_2^{179+}$ 
molecular ions at $R= 2$, $R= 2/90$, respectively. 
$^{a}$,$^{b}$, $^{c}$ The value are obtained with 
$\alpha^{-1}= 137.0359895$, $137.035999679$, $137.036$, respectively.
$^{d}$ Minmax result of ref. \cite{Fillion:2012}.}\label{tab:other} 
\end{table}

In table \ref{tab:other} we present a comparison with result from the 
literature.
It is obvious that the accuracy in these results is not high, but we 
must point out that these results are obtained with less computational 
effort than the present work, moreover, the result of ref. 
\cite{Fillion:2012} is obtained using minmax method. 
The results presented in the present work serve as a benchmark 
for the further development of methods in the relativistic domain.
Another issue is the accuracy regarding the 
precision of $\alpha$  value, where the comparison of our result in 
table \ref{tab:comp}, \ref{tab:other}, shows, as expected, that using 
$\alpha^{-1}_{old}=137.0359895$ (accuracy on the order $\sim 10^{-6}$ 
leads to uncertainty in the energy of the order $\Delta E \sim 10^{-12}$, 
i.e. it is on the order $(\Delta \alpha)^{2}$. 
The uncertainty in the $\alpha_{New}$ (CODATA18) used in the present 
work is of the order $1.5 \, 10^{-10}$ (see pml.nist.gov) leading to 
uncertainty of order $10^{-20}$ in the obtained energies. 
For $\rm H_{2}^{+} $, this uncertainty is larger than the precision 
reached in the calculation (this work), but it is smaller for 
${\rm Th}_2^{179+}$ as can be seen from table \ref{tab:Fres}, 
\ref{tab:comp}, where the uncertainty of the result for is of order 
$\sim 10^{-18}$. 
In addition, the accuracies of the values of ${\rm Th_{2}^{179+}}$ 
given in table \ref{tab:other} are blow the precision of 
$(\Delta\alpha^{-1}_{old})^{2}= 10^{-12}$.

Finally, in general, the point-like nucleus approximation is most 
commonly used, which is justified by the large scale difference 
between the nuclear radii and the internuclear distance.
The effect of finite nuclear size (FNS) for low-Z systems is expected 
to be small (or beyond interests), while it is significant for 
high-Z systems \cite{Artemyev:2010,Valuev:2020}.
The point-like model is typically used to compare different methods 
and the accuracies achieved.

In our work the higher order of the singular coordinate 
transformation eq. \ref{eq:transf} largely reduces the singular 
error as shown in Figs. \ref{fig:H2nu}, \ref{fig:Th2nu} and we would 
expect that the FNS effect on the relativistic effect is scaled down.
Corrections to the values of the physical quantities induced by the 
FNS are small, see for example
\cite{Karshenboim:2005,Athanasakis:2024,Alighanbari:2020,Valuev:2020},  
but they are relevant for comparing with the experiment. 
For ${\rm H}_2^+$ (and ${\rm HD}^+$) there is no particular need to 
improve the calculation of FNS corrections: The leading order is 
already known \cite{Korobov:2006,Aznabayev:2019} and higher-order 
corrections are too small to be of interest at the present level 
of accuracy. 
The FNS correction is important in the case of high-Z systems, 
especially when calculating QED corrections
\cite{Korobov:2021,Artemyev:2022,Artemyev:2015}.

For the calculations in the present work, we used quadruple precision.
The extension of the arithmetic is not required, rather parallelization 
of our code, which facilitate the computation and also allows 
higher grid points to be treated than those used in the tables 
\ref{tab:D40Nu8}, \ref{tab:D035Nu10}.
{Furthermore, multi-electron systems could be easily treated in the 
framework of Hartree-Fock or density functional theory} 
\cite{Kullie:1999,Kullie:2008}.   
Concerning the time of the calculation, it depends much on different 
parameter and the desired accuracy (and various grid sequences of 
different orders), which also influenced by the number of iteration in 
each grids, in particular for grids with large grid points.
For an example the calculation in table \ref{tab:D40Nu8} take about 
173 hours for relativistic calulation and 120 hours for nonrelatvistic  calulation, 
on one core (as already mentioned the code is not parallelized). 
For the same grid sequence the calculation in table \ref{tab:D035Nu10} 
takes about 2.5 times longer for ${\rm Th}_2^{179+}$, which is mainly 
because more iterations are required in the large grids. 
However, the last three grids in the grids-sequence take $60 \%$  of 
the time, although the gain in the accuracy is only about two or three 
orders.      
\subparagraph{Conclusion and Outlook}
In this work, we presented highly accurate minimax FEM calculations 
for the ionic molecule ${\rm H}_2^+$ and quasi-molecule ${\rm Th}_2^{179+}$. 
We showed systematically accurate values by investigation of the 
convergence behavior of the relativistic and non-relativistic numerical  
solution of the two-center Coulomb problem. 
Our result is compared with results available in the literature and 
shows a good agreement with the recently published result. 
Applications such as the g-factor (tensor) of bound electrons for 
${\rm H}_2^+$ by perturbative evaluation of the Zeeman energy
\cite{Hegstrom:1979,Karr:2021,Korobov:2021,Pachucki:2023,Beier:2000} 
are currently being investigated, taking into account relevant 
corrections for comparison with the experimental value. 

The high-precision relativistic calculations we achieved enables
investigation of other properties such as QED corrections.
However, the calculations of radiative QED corrections are demanding 
and   necessitate sums over intermediate states\cite{Fillion:2012}. 
Artemyev et al \cite{Artemyev:2022} have done some calculations on 
heavy one-electron quasi-molecular ion such as ${\rm U}_2^{183+}$, 
the precision they achieved is not high.
The most attractive application is the one-loop self-energy, which is 
currently the main source of theoretical uncertainty in the hydrogen 
molecular ions \cite{Korobov:2021,Korobov:2017}. 
The calculation of the  one-loop self-energy in a weak binding 
field (i.e. low nuclear charges), suffers from a serious loss of 
numerical precision because of strong cancellations occurring in the 
renormalization procedure, hence the need for extremely accurate wave 
functions and energies \cite{Jentschura:1999,Jentschura:2001}.
Solving the two-center Dirac equation with FEM-minmax, offers the 
possibility to improve the precision substantially, especially that 
the minmax solution is free from spurious states and contamination 
of the negative (positronic) continuum, as already mentioned. 
\subparagraph{Acknowledgments} 
I would like to thank Prof. M. Garcia  (University Kassel),  
Prof. S. Schiller (University D\"usseldorf) for their supports. 
And  Dr. Jean-Philippe Karr 
(Universit\`{e} d{'}\`{E}vry-Val-d{`}\`{E}ssonne) 
and Dr. Anton N. Artemyev (University of Kassel) and Prof. D. Kolb 
(University of Kassel, retired) for valuable discussion. 
I also thank the computing center of the University D\"usseldorf for 
providing resources and advice (FEMCalculation Project). 
I am grateful to the anonymous reviewer for his constructive and 
valuable comments to improve the manuscript.
\appendix 
\section{}\label{AA}
In this appendix we briefly put forward some materials that can help 
the reader follow the discussion in Sec. \ref{RaD}. For details see 
\cite{Kullie:20042,Kullie:2001,Kullie:2003}.
\subparagraph{A 1.}\label{A1}
The axial symmetry around the internuclear axis (the $z$-axis) in 
two-center case suggests to use of the well-known prolate spheroidal 
(elliptic spheroidal) coordinates $\xi, \,\eta, \,\varphi$,
\begin{eqnarray}\nonumber
&&x=\frac{R}{2} u(\xi,\eta) \, \cos\varphi, \,y=\frac{R}{2} u(\xi,\eta) \, 
\sin\varphi, z=\frac{R}{2}\xi\cdot \eta, 
\\\nonumber
&&\mbox{where } u(\xi,\eta)=
\sqrt{(\xi^{2}-1)(1-\eta^{2})}\,\
\end{eqnarray}
where $R$ is the inter-nuclear distance in atomic units and  $\varphi$  
is the electron's angular coordinate. 
The  angular coordinate is separable and the problem is reduced to 
a 2-dimensional one.   
The distances of the electron to the nuclei are given by 
$r_{1}=(\xi+\eta) \frac{R}{2},\, r_{2}=(\xi-\eta) \frac{R}{2}.$
The Coulomb singularity of point nucleus model causes a singular 
behavior of the relativistic solutions at the position of the nuclei 
of the form $r^{-1+\gamma_{l,\kappa}}_{l}$, with  
$\gamma_{l,\kappa}=\sqrt{\kappa^{2}-{(\alpha Z_{l})^2}}$ and 
$|\kappa|=|j_{z}|+\frac{1}{2},\,l=1,2$, it is well-known from atomic 
calculations \cite{Yang:1993,Duesterhoeft:1994}. 
Thus, further singular coordinate transformation is needed 
\cite{Kullie:2001,Kullie:20042,Kullie:2003} as the following, 
\begin{eqnarray}\label{eq:transf} 
 Y  & = &1+\sum_{i=1}^{\nu/2} d_{i}\, S^{\nu+2(i-1)}(x/2)
 \\\nonumber
  & & \mbox{for } \nu=2,4,6,8, 10
\end{eqnarray}
 where $Y$ stands for $\xi$ or $\eta$ and  $S$ for $\sinh$ or 
 $\sin$ and $x$ for $s$ or $t$, respectively.
 With $0 \le s < \infty,  \, 0 \le t \le \pi$.  
The transformation can be calculated by integration of the following 
derivatives,
\[\frac{d Y}{dx}=D_n\, S^{2n+1}(x),\,  D_n=\frac{(2n+1)!}{n!\,2^{n}}, 
\mbox{ with }  n=\frac{\nu}{2}-1\]
which determines the coefficients $d_i$ in eq \ref{eq:transf}.
Mathematically they are connecting to the hyper-geometric function 
$_2F^{1}$ \cite{Kullie:20042}, which can be found by performing the 
integration using e.g. {\sl  Mathematica}. 
The transformation regularizes the singularities at the nuclei by 
increasing the point density in the inner region. 
The higher $\nu$, the denser the points near the nuclei, which ensures  
a better approximation of (the singularity of) the wave function  
\cite{Kullie:2003}. 
An advantage of this transformation is the use of a square grid type 
over $s$ and $t$, since 
$\lim_{s\to 0}(\xi-1)\sim s^{\nu}, \lim_{t\to, 0,\pi}(1-\eta) 
\sim t^{\nu}$.
\subparagraph{A 2.}\label{A2}
In the present FEM treatment, the 2-dimensional domain $(s,t)$ is 
subdivided into triangular elements $\rm e$.
The component ${k}$ of the relativistic wave function is approximated by 
\begin{eqnarray}\label{glob1}
\phi^{k}(s,t)= G^{{k}}(s,t) \sum_{{\rm e}} 
\sum_{i}^n d^{k,\rm e}_i N_{i}^{{k,\rm e}}(s,t)\ ,
\end{eqnarray}
where $\rm e$ is the number of the element, $n$ is the total number of 
the nodal points of the element $\rm e$ and $d^{k,\rm e}_i$ are the 
unknown coefficients at the nodal points $i$. 
The shape functions $N_i^{\rm e}(s,t)$ are defined inside the element 
$\rm e$ by complete polynomials of an order $p$ in $s,t$ and zero 
elsewhere \cite{Heinemann:1987}. 
The functions $G^{k}(s,t)$ account for the global behavior of the wave 
function.
They are given by 
\begin{eqnarray}\label{glob02}
G^{{k}}(s,t)&=& G^{{k}}_{1}(s,t) \cdot G_{2}(s,t),\nonumber\\ 
G_1^{{k}}(s,t)&=& ((\xi^{2}-1)(1-\eta^{2}))^{\frac{m_k}{2}}, \\
\nonumber
\label{glob2}
G_2(s,t)&=& r_{1}^{-1+\gamma_{1,\kappa}}\cdot r_{2}^{-1+\gamma_{2,\kappa}},
\\\nonumber
m_{k}&=&j_z+(-1)^{k}/2\ . 
\end{eqnarray} 
where $G_1^{k}(s,t)$ represents the angular momentum dependence and  
$G_2(s,t)$ expresses the singular behavior at the two nuclei, and 
$\gamma_{l,\kappa}$ as given above. 
The larger $Z$, the  smaller  $\gamma_{l,\kappa}$ and the singular 
behavior of the wave function is stronger, hence the convergence 
becomes less efficient. 
\providecommand{\noopsort}[1]{}\providecommand{\singleletter}[1]{#1}%
%
\end{document}